\documentclass[aps,prl,twocolumn,superscriptaddress]{revtex4-1}
\pdfoutput=1

\usepackage{amsmath}
\usepackage{amsfonts}
\usepackage{amssymb}
\usepackage{bm}
\usepackage{graphicx}

\begin{document}


\title{Atomic energy mapping of neural network potential}



\author{Dongsun Yoo}
\thanks{These three authors contributed equally.}

\author{Kyuhyun Lee}
\thanks{These three authors contributed equally.}

\author{Wonseok Jeong}
\thanks{These three authors contributed equally.}
\affiliation{Department of Materials Science and Engineering and Research Institute of Advanced Materials, Seoul National University, Seoul 08826, Korea}

\author{Satoshi Watanabe}
\affiliation{Department of Materials Engineering, The University of Tokyo, Bunkyo, Tokyo 113-8656, Japan}

\author{Seungwu Han}
\email[]{hansw@snu.ac.kr}
\affiliation{Department of Materials Science and Engineering and Research Institute of Advanced Materials, Seoul National University, Seoul 08826, Korea}


\date{\today}

\begin{abstract}
We show that the intelligence of the machine-learning potential arises from its ability to infer the reference atomic-energy function from a given set of total energies. By utilizing invariant points in the feature space at which the atomic energy has a fixed reference value, we examine the atomic energy mapping of neural network potentials. Through a series of examples on Si, we demonstrate that the neural network potential is vulnerable to `ad hoc' mapping in which the total energy appears to be trained accurately while the atomic energy mapping is incorrect in spite of its capability. We show that the energy mapping can be improved by choosing the training set carefully and monitoring the atomic energy at the invariant points during the training procedure.
\end{abstract}


\maketitle


Recently, machine-learning (ML) approaches to developing interatomic potentials are attracting considerable attention because it is poised to overcome the major shortcoming inherent to the classical potential and first-principles method, i.e., difficulty in potential development and huge computational cost, respectively. Favored ML models are the neural network~\cite{14-Behler2007,DeepMD} and Gaussian process~\cite{20-GAP}. In particular, the high-dimensional neural network potential (NNP) suggested by Behler and Parrinello\cite{14-Behler2007} is attracting wide interests with applications demonstrated over various materials encompassing metals,\cite{15-Artrith2014,16-Boes2017,22-Eshet2012} insulators,\cite{17-Artrith2016,Watanabe2017} semiconductors,\cite{18-GeTe,21-BehlerSi} and molecular clusters.\cite{19-Kolb2016}


While the methodological advances are under rapid progress~\cite{5-wACSF,6-LiGA,7-CUR,8-StratifiedNN,9-ImplantedNN,10-MixtureModel,11-GDF}, the conceptual foundation of NNP is still elusive, partly due to the black-box nature of the neural network. Furthermore, NNP infers atomic energies while it is trained over total energies that are sums of atomic energies. This obscures the nature of training procedure and makes it difficult to assess learning quality. Responding to this, in this Letter, we try to address basic questions on NNP such as `what is the intelligence of NNP?' and `how the learning quality is determined?'. At variance with general views, we show that the core of training procedure in NNP is to infer the reference atomic energy grounded on the density functional theory (DFT), from the given relationship between the structure and total energy. With examples on Si, we demonstrate that  NNP is prone to ad hoc mapping in which the total energy is trained accurately but the atomic energy mapping is incorrect. We also show that the reference atomic energies can serve as a tool to assess the learning quality of NNP.

Most ML potentials are based on the representability of the DFT total energy ($E_\text{tot}^\text{DFT}$) as a sum of the atomic energy ($E_\text{at}$) that depends on the local environment within a certain cutoff radius ($R_\text{c}$):
\begin{equation}\label{eq:1}
E_\text{tot}^\text{DFT} = \sum_i {E_\text{at}(\mathcal{R}_i;R_\text{c})}\,,
\end{equation}
where $i$ is the atom index and $\mathcal{R}_i$ is the collection of relative position vectors of atoms lying within $R_\text{c}$ from the $i$th atom. (For simplicity, we assume a unary system that is large enough that various cutoff spheres in the following discussions do not self-overlap under periodic boundary conditions and wave functions are effectively real-valued.) As is well known, the total energy can be expressed by integration of the local energy density, although it is not unique.\cite{[{}] [{, Appendix H and references therein.}]24-MartinBook} Then, by partitioning the space into non-overlapping atomic volumes, one can assign energies to each atom whose sum equals to the total energy.\cite{25-Yu2011,26-QCTFF} In consideration of locality or `nearsightedness' of the electronic structure,\cite{1-Kohn1996} which empowers the $\mathcal{O}(N)$ approach \cite{2-Goedecker1999}, it would be formally viable to define $E_\text{at}(\mathcal{R}_i;R_\text{c})$ in Eq. \eqref{eq:1} within DFT, which depends only on local features and so is transferable. In the below, we elaborate on this explicitly, with a particular attention to the transferable range.

Within the semilocal density approximation, $E_\text{tot}^\text{DFT}$ can be expressed in terms of the one-electron density matrix $\rho(\mathbf{r},\mathbf{r'})$ and the electron density $\rho(\mathbf{r})=\rho(\mathbf{r},\mathbf{r})$:
\begin{equation}\label{eq:3}
\begin{split}
&E_\text{tot}^\text{DFT} = E_\text{kin} + E_\text{XC} + E_\text{Coul}\\
&=-\frac{1}{2}\int\nabla_{\bf r}^2 \rho(\mathbf{r},\mathbf{r'}) \rvert_{\mathbf{r}=\mathbf{r}'} d\mathbf{r}' + \int \rho(\mathbf{r}) \varepsilon_\text{XC}(\rho(\mathbf{r}),\nabla \rho(\mathbf{r})) d\mathbf{r}\\
&+\frac{1}{2} \int \frac{\rho(\mathbf{r})\rho(\mathbf{r}')}{\lvert \mathbf{r}-\mathbf{r}' \rvert} d\mathbf{r}d\mathbf{r}' - \sum_i \int \frac{q_i \rho(\mathbf{r})}{\lvert \mathbf{r}-\mathbf{r}_i \rvert} d\mathbf{r} + \sum_{i>j} \frac{q_i q_j}{\lvert \mathbf{r}_i-\mathbf{r}_j \rvert}\,,
\end{split}
\end{equation}
where the atomic unit is used, $\varepsilon_\text{XC}$ is the exchange-correlation energy density, $q_i$ and $\mathbf{r}_i$ are the ionic charge and position of the $i$th atom, respectively. Under the assumption that $\mathcal{O}(N)$ methods, in particular the divide-and-conquer (DAC) approach\cite{3-Yang1991,4-Yang1995}, work well for given systems, we will explicitly show that i) each energy term can be split without any loss into atomic contributions that are defined locally around each atomic site, and ii) the atomic energy depends only on nearby atoms such that it is transferable to other systems as long as local environments are maintained.  

We start with partitioning the space into atomic cells without gaps or overlapping (for instance, Voronoi cells). Let $V_i$ be the cell enclosing the $i$th atom. We define $\rho_i(\mathbf{r})$ as
$\rho_i(\mathbf{r}) = \rho(\mathbf{r})\lbrack\mathbf{r} \in V_i \rbrack\,$
where [...] is the Iverson bracket whose value is 1 (0) when the logical proposition in the bracket is true (false). Obviously, $\rho(\mathbf{r})=\sum_i \rho_i(\mathbf{r})$. It is easily seen that $E_\text{XC}$ is the sum of the atomic exchange-correlation energy ($E_{\text{XC},i}$) that is obtained by substituting $\rho_i(\mathbf{r})$ for $\rho(\mathbf{r})$ in the integrand of $E_\text{XC}$. As is assumed in the DAC method,\cite{3-Yang1991} the charge density at a certain point is influenced by only nearby atoms if the local chemical potential of electrons is fixed. This means that  $\rho_i(\mathbf{r})$, and hence $E_{\text{XC},i}$ is affected by atomic arrangements within a certain cutoff ($R_\text{c}^1$) from $\mathbf{r}_i$. 

Next, we define the total charge density in $V_i$: $\rho_{\text{tot},i}(\mathbf{r}) = q_i\delta(\mathbf{r}-\mathbf{r}_i) - \rho_i(\mathbf{r})$. 
It is straightforward to show that $E_\text{Coul}$ can be expressed as a summation of the atomic Coulomb energy, $E_{\text{Coul},i}$, defined as follows:
\begin{multline}\label{eq:6}
E_{\text{Coul},i} = \frac{1}{2}\sum_{j \neq i} \int \frac{\rho_{\text{tot},i}(\mathbf{r})\rho_{\text{tot},j}(\mathbf{r}')}{\lvert \mathbf{r}-\mathbf{r}' \rvert} d\mathbf{r} d\mathbf{r}'\\
+ \frac{1}{2} \int \frac{\rho_i(\mathbf{r})\rho_i(\mathbf{r}')}{\lvert \mathbf{r}-\mathbf{r}' \rvert} d\mathbf{r} d\mathbf{r}'- \int \frac{q_i \rho_i(\mathbf{r})}{\lvert \mathbf{r}-\mathbf{r}_i \rvert} d\mathbf{r}.
\end{multline}
The first term on the right-hand side of Eq.~(\ref{eq:6}) is long-ranged, which is incompatible with the finite cutoff. However, electrostatic interactions are effectively screened or cancelled in many condensed phases, so it would be a reasonable approximation to ignore them beyond a certain cutoff ($R_\text{c}^2$). Thus, we omit the Coulomb interaction between $\rho_{\text{tot},i}$ and $\rho_{\text{tot},j}$ if $\lvert \mathbf{r}_j-\mathbf{r}_i \rvert > R_\text{c}^2$. Since $\rho_i(\mathbf{r})$ and $\rho_{\text{tot},i}(\mathbf{r})$ are influenced by atoms within $R_\text{c}^1$ (see above), $E_{\text{Coul},i}$ depends on atoms inside $R_\text{c}^1 + R_\text{c}^2$ (neglecting the volume of $V_i$). To note, some implementations of NNP explicitly describe the long-range Coulomb potential, separately from short-ranged atomic energies.\cite{28-ArtrithZnO,29-WaterDimer}

As the last step, we discuss the locality of $E_\text{kin}$. Since $\rho(\mathbf{r},\mathbf{r}')$ decays exponentially  with $\lvert \mathbf{r} - \mathbf{r}' \rvert$ in insulators and metals at finite temperatures~\cite{2-Goedecker1999}, one can neglect $\rho(\mathbf{r},\mathbf{r}')$ when $\lvert \mathbf{r} - \mathbf{r}' \rvert$ is bigger than a cutoff ($R_\text{c}^3$), which is utilized in the density-matrix-based DAC method~\cite{4-Yang1995}. Therefore, for a given position $\mathbf{r}$, $\rho(\mathbf{r},\mathbf{r}')$ is determined by the atomic configurations within a cutoff distance ($R_\text{c}^4$) from $\mathbf{r}$, which should be larger than $R_\text{c}^3$. With the projected density matrix $\rho_{ij}(\mathbf{r},\mathbf{r}')=\rho(\mathbf{r},\mathbf{r}')\lbrack \mathbf{r} \in V_i \rbrack \lbrack \mathbf{r}' \in V_j \rbrack$, we define the atomic density matrix $\rho_{\text{at},i}(\mathbf{r},\mathbf{r}')$ as follows:
\begin{equation}\label{eq:7}
\rho_{\text{at},i}(\mathbf{r},\mathbf{r}')=\rho_{ii}(\mathbf{r},\mathbf{r}') + \frac{1}{2}\sum_{j \neq i}^{\lvert \mathbf{r}_j - \mathbf{r}_i \rvert < R_\text{c}^3} \rho_{ij}(\mathbf{r},\mathbf{r}')\,.
\end{equation}
It can be shown that $\rho(\mathbf{r},\mathbf{r}')=\sum_i \rho_{\text{at},i}(\mathbf{r},\mathbf{r}')$ and $\rho_{\text{at},i}(\mathbf{r},\mathbf{r}')$ depends only on the atomic arrangements within $R_\text{c}^4$ from the $i$th atom (neglecting the volume of $V_i$). The atomic kinetic energy is then given in the following:
\begin{equation}\label{eq:8}
E_{\text{kin},i}=-\frac{1}{2}\int\nabla_{\bf r}^2 \rho_{\text{at},i} (\mathbf{r},\mathbf{r}')\rvert_{\mathbf{r}=\mathbf{r}'} d\mathbf{r}' .
\end{equation}
Since the kinetic-energy operator is linear, the sum of the atomic kinetic energy is equivalent to the total kinetic energy.

Combining the above analyses, the atomic energy of the $i$th atom formally derives from the DFT calculations:
\begin{equation}\label{eq:9}
E_{\text{at},i}=E_{\text{kin},i}+E_{\text{XC},i}+E_{\text{Coul},i},
\end{equation}
and $E_\text{tot}^\text{DFT} = \sum_i E_{\text{at},i}$.
By evaluating $E_{\text{at},i}$ in various structures, one can obtain in principle the atomic energy as a continuous function of the local environment:
\begin{equation}\label{eq:10}
E_{\text{at},i} \longrightarrow E_\text{at}^\text{DFT}(\mathcal{R};R_\text{c})\,,
\end{equation}
where $R_\text{c} = \max(R_\text{c}^1+R_\text{c}^2, R_\text{c}^4)$.
Note that the atomic energy is not unique because it depends on the way to define atomic cells.

The existence of $E_\text{at}^\text{DFT}$ implies that the objective of the present machine learning is to identify  $E_\text{at}^\text{DFT}$ when only total energies are informed. This is at variance with the conventional view that NNP is merely an interpolation of given total energies.~\cite{Angewandte,Amp} Mathematically, the neural network has the capability to infer the underlying function when only sums of function values are provided. (See examples in Supplemental Material~\footnote[1]{See Supplemental Material below for the capability of neural network learning from the sum, atomic energy mapping of Ni nanocluster, the training set and computational details, the estimation of prediction uncertainty, the analysis on the distance to training set, and the metric to quantify connectivity in $\mathbf{G}$ space.} on a piecewise cubic spline and the embedded atom potential.) To reduce the huge dimension of $\mathcal{R}$ and obtain $E_\text{at}$ in a computationally feasible way, two approximations are adopted. First, the cutoff radius is reduced from $R_\text{c}$, which should be fairly large for high accuracy, to $r_\text{c}$ that is usually chosen to be 6-7 $\text{\AA}$. This is a reasonable range because the chemical influence rapidly diminishes beyond this boundary. Second, the local environment is described by feature vectors whose dimension is significantly lower than for $\mathcal{R}$. The popular choices are smooth-overlap-of-atomic-positions (SOAP)~\cite{SOAP} or symmetry function  vectors ($\mathbf{G}$)~\cite{GG}. These feature vectors also automatically incorporates rotational and translational invariance inherent to the atomic energy. Here, we employ the symmetry function. Thus, 
\begin{equation}\label{eq:11}
E_\text{tot}^\text{DFT} = \sum_i {E_\text{at}^\text{DFT}(\mathcal{R}_i;R_\text{c})}  \simeq \sum_i E_\text{at}^\text{NN}(\mathbf{G}_i;r_\text{c})\,.
\end{equation}

The accuracy of NNP therefore hinges on how close $E_\text{at}^\text{NN}$ obtained through machine learning is to the reference $E_\text{at}^\text{DFT}$ over the configurational space spanned by the given training set. However, since $E_\text{at}^\text{NN}$ is fitted to the total energies, rather than directly to $E_\text{at}^\text{DFT}$, the ML procedure does not necessarily guarantee sufficient accuracies in $E_\text{at}^\text{NN}$. That is to say, $E_\text{at}^\text{NN}$ can reproduce total energies in the training set precisely but deviate significantly from $E_\text{at}^\text{DFT}$. Indeed, we will demonstrate that NNP is vulnerable to such `ad hoc' energy mapping, which leads to incorrect total energies in related configurations and undermines the transferability of NNP. To note, the ad hoc mapping should be distinguished from the gauge-dependent degree of freedom in the energy density.\cite{24-MartinBook} 

Even though the existence of $E_\text{at}^\text{DFT}$ was shown formally in the above, the actual calculation of $E_\text{at}^\text{DFT}$ would be highly costive. (We  note a recent effort to directly train NNP over atomic energies.~\cite{27-AtomENNP}) Furthermore, there exist an infinite number of valid $E_\text{at}^\text{DFT}$, making it hard to grade the energy mapping of $E_\text{at}^\text{NN}$. However, there are invariant  points in the $\mathbf{G}$ space at which $E_\text{at}^\text{DFT}$ is uniquely defined without any degree of freedom. For instance, in the crystalline Si, all the atoms are equivalent, and so the total energy per atom is simply equal to $E_\text{at}^\text{DFT}$ for the corresponding $\mathbf{G}$. Transforming lattice vectors of the unit cell also results in similar conditions. In the below, utilizing these special $\mathbf{G}$ points, we will analyze the atomic energy mapping for three examples on Si that are progressively more complicated.

\begin{figure}
	\includegraphics{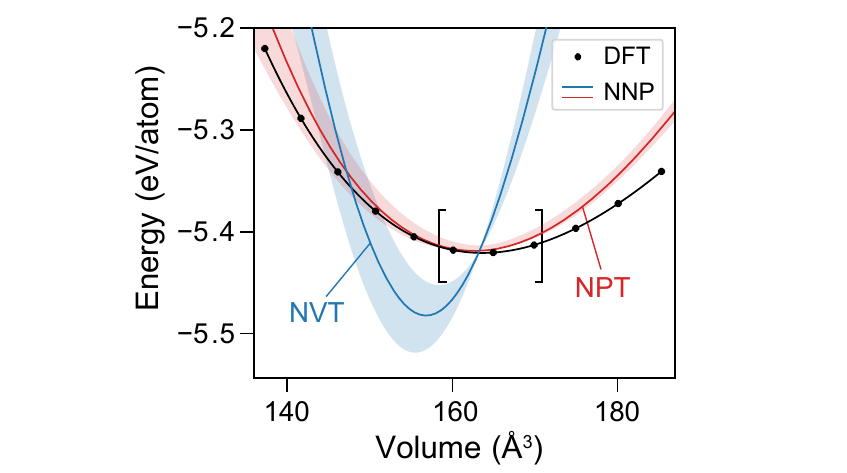}
	\caption{The equation of state (EOS) for Si crystal compared between DFT and NNP. The blue and red solid lines are the average EOS over five NNPs that are trained with NVT- and NPT-MD snapshots, respectively. The shades are one standard-deviation from the average, corresponding to the prediction uncertainty. The squared bracket indicates the volume range where corresponding $\mathbf{G}$'s lie in the proximity of the training set.\label{fig:1}}
\end{figure}

As the first example, we train NNPs for crystalline Si by adopting an in-house code named SIMPLE-NN\cite{12-SIMPLENN}. The training (validation) set consists of 350 (150) MD snapshots of the 64-atom cubic supercell under the NVT condition of 1000 K and the equilibrium volume at 0 K. After training, the root-mean-squared error (RMSE) in the total energy and atomic force is 0.9 (0.9) meV/atom and 0.10 (0.11) eV/$\text{\AA}$ for the training (validation) set, respectively. (See the Supplemental Material~\cite{Note1} for further details in DFT calculations and NNP training.) Atomic vibrations during MD give rise to local expansion or compression. As a result, atomic configurations around certain Si atoms resemble those in the crystalline phase under hydrostatic pressures, which forms the equation of state (EOS) and corresponds to invariant $\mathbf{G}$ points explained above. This is confirmed by principal-component analysis (PCA) and measuring the shortest distances from the invariant $\mathbf{G}$ points to the training set. (See Supplemental Material~\cite{Note1}.) This implies that atomic energies at the $\mathbf{G}$ points along EOS are learnable although they do not belong to the training set. Therefore, if atomic energies are properly mapped, NNP should be able to predict correctly the energy-volume relation at 0 K.

Figure \ref{fig:1} compares EOS inferred by the as-trained NNP (blue line) with DFT results (black dots). The light shade means prediction uncertainty evaluated by ensembles of NNP~\cite{13-AddressingUncertainty} (see also Supplemental Material~\cite{Note1}). The squared bracket indicates the range of the volume whose $\mathbf{G}$ is in close proximity to the training set. Interestingly, NNP predicts correctly the energy at equilibrium but energies at other volumes significantly deviate from the DFT curve with errors far bigger than RMSE in the total energy. That is to say, NNP predicts the total energy correctly but the atomic energy is markedly wrong, which corresponds to the ad hoc energy mapping. From the continuity in $E_\text{at}$, the incorrect energy mapping should affect other training points neighboring invariant $\mathbf{G}$ points, implying that the ad hoc mapping extends over a significant portion of the training set.

The ad hoc mapping in the above example happens because the training set consists of structures with a fixed volume. This condition constrains the local expansion and contraction to occur concurrently within the same structure. Consequently, any additional atomic energy that varies linearly with the volume does not affect the total energy, and so the slope of EOS at the equilibrium volume becomes an arbitrary number. The ad hoc mapping in this case can be resolved by considering structures with different volumes or including virial stress in the loss function. For instance, the red line in Fig. \ref{fig:1} shows EOS predicted with NNPs that are trained with MD snapshots from NPT ensembles at 1000 K and zero pressure. During MD, the supercell expands or shrinks, avoiding the exact cancellation among the local volume changes. As a result, it is seen that NNPs can predict the slope and curvature of EOS reasonably.
 
\begin{figure}
	\includegraphics{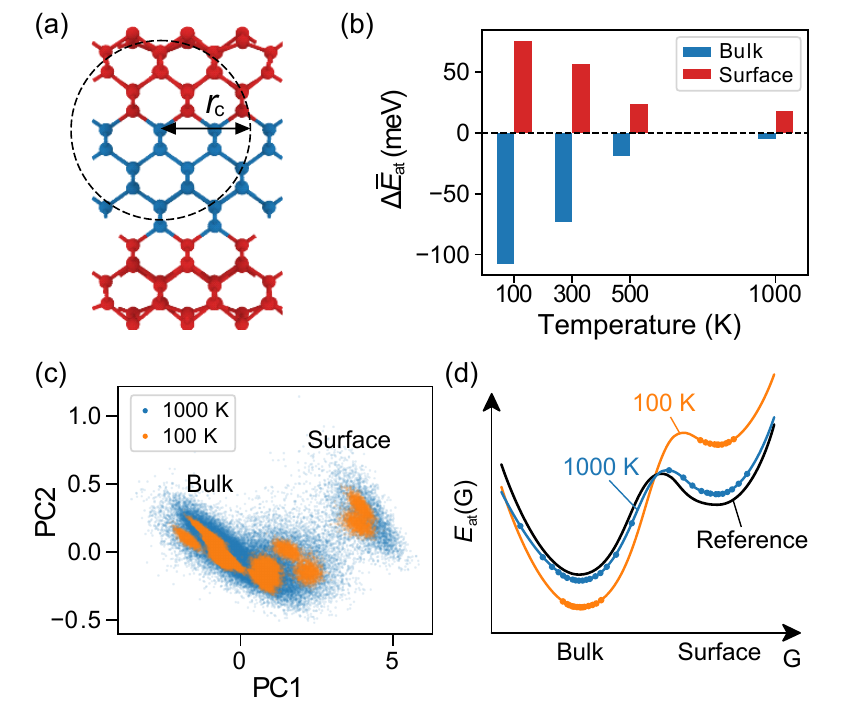}
 	\caption{(a) The structure of Si(100)-(2$\times$2) slab. The atoms in bulk and surface regions are marked in blue and red, respectively. $r_\text{c}$ is the cutoff radius of symmetry functions. (b) The average of atomic-energy difference between DFT and NNPs for bulk and surface groups, plotted against the temperature of the training set. (c) Scatter plot along principal components (PC) of $\mathbf{G}$ vectors in the training set. (d) Schematic illustration of ad hoc mapping due to separate groups of training points.\label{fig:2}}
\end{figure}

The second example concerns a surface model of Si. The training set consists of MD trajectories of Si(100)-(2$\times$2) symmetric slab in the NVT condition at a certain temperature between 100 and 1000 K [see Fig. \ref{fig:2}(a)]. To assess the learning quality, we compare atomic energies for the geometry relaxed at 0 K with DFT. Unlike crystalline Si in the previous example, the reference $E_\text{at}^\text{DFT}$ is not available directly. Nevertheless, Si atoms inside the slab (blue atoms) have neighborhood similar to that in the crystal (see a dashed circle). Therefore, $E_\text{at}$ in this region should be close to the crystalline $E_\text{at}^\text{DFT}$ at the equilibrium volume [$E_\text{at}^\text{DFT}(\text{bulk})$]. Since $E_\text{tot}^\text{DFT}$ is available for the whole structure, the average $E_\text{at}^\text{DFT}$ for the surface region (red atoms) can be obtained as $[E_\text{tot}^\text{DFT} - N_\text{b} \cdot E_\text{at}^\text{DFT}(\text{bulk})]/N_\text{s}$, where $N_\text{b}$ and $N_\text{s}$ are the number of atoms in the bulk and surface regions, respectively. By taking the difference in averaged values of $E_\text{at}^\text{NN}$ and ${E}_\text{at}^\text{DFT}$ in each region, one can quantify average mapping errors, $\Delta \bar{E}_\text{at}(\text{bulk})$ and $\Delta \bar{E}_\text{at}(\text{surface})$, respectively.

Figure \ref{fig:2}(b) presents $\Delta \bar{E}_\text{at}(\text{bulk})$ and $\Delta \bar{E}_\text{at}(\text{surface})$ for NNPs trained over MD trajectories at different temperatures. At a low temperature of 100 K, the mapping error is $-108$ and 76 meV/atom for bulk and surface regions, respectively, which is far bigger than RMSE (0.3 meV/atom). This is another example of ad hoc energy mapping; NNP correctly predicts the total energy because errors in the atomic energy mapping cancel with each other. In Fig. \ref{fig:2}(b), it is intriguing that the mapping error gradually decreases as the temperature in the training set increases, and at the high temperature of 1000 K, the magnitude of mapping errors becomes comparable to RMSE in the total energy (3 meV/atom).

To understand the temperature-dependent mapping error, we examine in Fig. \ref{fig:2}(c) the distribution of training points in the $\mathbf{G}$ space using PCA on the training sets at 100 and 1000 K. It is seen that at 100 K, the training points corresponding to the bulk and surface region are well separated. In contrast, energetic vibrations at 1000 K result in much broader distribution of training points such that bulk and surface regions are slightly connected. (Other combinations of principal axes show similar behaviors.) As schematically drawn in Fig. \ref{fig:2}(d), if  clusters of training points are separate as in 100 K, the machine learning is prone to ad hoc mapping because any cancelling offsets give almost the same total energy and atomic forces. On the other hand, at higher temperatures with every region  connected to some degrees, $E_\text{at}$ at intermediate configurations helps adjust the energy offset between basins. In Supplemental Material~\cite{Note1}, we define a metric that can quantify the connectivity in the $\mathbf{G}$ space and show that the mapping error is sufficiently small when the connectivity is higher than a cutoff. 


\begin{figure}
	\includegraphics{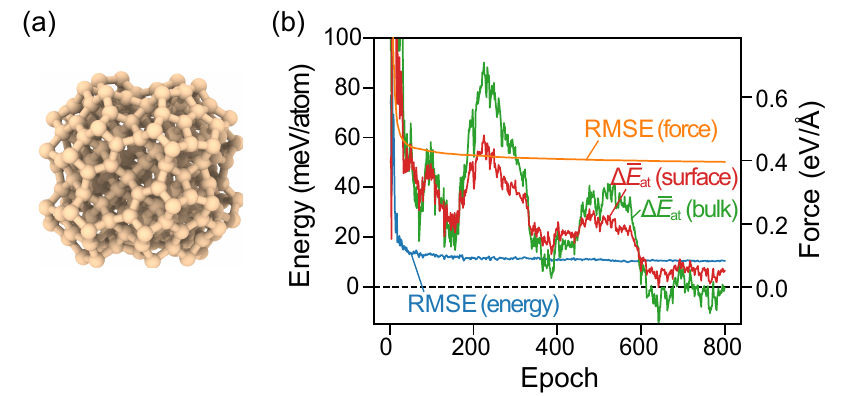}
	\caption{(a) Si$_{239}$ nanocluster relaxed at 0 K. (b) Change of RMSE for energy and force, and mapping errors for surface and bulk regions in Si(100)-(2$\times$2) slab in Fig. \ref{fig:2}(a), with respect to the training epoch.\label{fig:4}}
\end{figure}

Albeit simple, the above cases substantiate the ad hoc mapping that originates from limitations in the training set. In practice, a single training set usually encompasses diverse structures such as bulk, surfaces, and defects, and chances are that the ad hoc mapping can be avoided in principle. Nevertheless, the error-cancelling energy offsets as in Fig. \ref{fig:2}(d) are omnipresent, which can go unnoticed if the training procedure is monitored by RMSE only. To show this, we generate a training set from MD simulations of a 239-atom Si nanocluster with Wulff-constructed \{100\}, \{110\}, and \{111\} facets at 1000-1700 K. [See Fig. \ref{fig:4}(a) for the structure relaxed at 0 K.] The analysis on the connectivity (see above) confirms that training points are well connected.

In Fig. \ref{fig:4}(b), we plot RMSE for the total energy and force with respect to the training epoch. It also shows $\Delta \bar{E}_\text{at}(\text{bulk})$ and $\Delta \bar{E}_\text{at}(\text{surface})$ for the (100)-(2$\times$2) surface model in Fig. \ref{fig:2}(a). The analysis similar to Fig. S3 shows that the $\mathbf{G}$ points in the (100)-(2$\times$2) slab model are in the vicinity of training points, and hence they are learnable. Therefore, NNP is expected to predict surface and bulk energies in reasonable agreement with DFT results. In Fig. \ref{fig:4}(b), it is seen that RMSE remains almost constant after about 100 epochs while $\Delta \bar{E}_\text{at}(\text{bulk})$ and $\Delta \bar{E}_\text{at}(\text{surface})$ converge at much slower rates. This indicates a risk in concluding the training convergence in terms of RMSE, and supports $E_\text{at}$ at invariant $\mathbf{G}$ points as  alternative convergence parameters. Obviously, if  the crystalline structures are included in the training set, $\Delta \bar{E}_\text{at}(\text{bulk})$ would converge as fast as RMSE, but this does not guarantee the proper energy mapping  at other training points. Therefore, we suggest to collect invariant $\mathbf{G}$ points as a separate test set for monitoring the atomic energy mapping, rather than including them in the training set, at least in the initial stage of training. 

After a sufficient number of epochs, the surface energies for (100)-(2$\times$2), (110)-(2$\times$1), and (111)-(2$\times$1) slab models that are fully relaxed by NNP agree with DFT results within 8\%.  (The corresponding errors by NNP trained up to 200 epochs are within 20\%.) It is intriguing that just one type of structure (nanocluster) can train NNP over such a wide range of configurations when the energy mapping is correct. This implies that NNPs with proper mapping are more transferable than those with ad hoc mapping, which may contribute to improving the stability of MD simulations.\cite{11-GDF} It will be also useful in developing general-purpose NNPs.\cite{30-PRXGeneralSi} Finally, we find that monitoring the energy mapping is helpful in selecting training parameters such as the regularization parameter of the neural network.

In conclusion, we showed that the aim of training NNP is to learn the atomic energy function defined at the DFT level from total energies, and the transferability of NNP lies in the accuracy of atomic energy mapping. The invariant $\mathbf{G}$ points with the unique ${E}_\text{at}^\text{DFT}$ provided ways to examine the atomic energy mapping. Several examples confirmed that NNP is vulnerable to ad hoc mapping due to limitations in the training set and/or certain choices of computational parameters. The energy mapping can be improved by choosing the training set carefully and monitoring the atomic energy at the invariant points during the training procedure.
 By clarifying what NNP actually learns, the present work will contribute to constructing accurate and transferable machine-learning potentials.

\begin{acknowledgments}
This work was supported by Technology Innovation Program (10052925)  by  Ministry of Trade, Industry \& Energy, and Creative Materials Discovery Program by the National Research Foundation (2017M3D1A1040689). The computations were carried out at the National Supercomputing Center (KSC-2018-CHA-0038).
\end{acknowledgments}

\bibliography{ref}

\end{document}



\title{Atomic energy mapping of neural network potential}



\author{Dongsun Yoo}
\thanks{These three authors contributed equally.}

\author{Kyuhyun Lee}
\thanks{These three authors contributed equally.}

\author{Wonseok Jeong}
\thanks{These three authors contributed equally.}
\affiliation{Department of Materials Science and Engineering and Research Institute of Advanced Materials, Seoul National University, Seoul 08826, Korea}

\author{Satoshi Watanabe}
\affiliation{Department of Materials Engineering, The University of Tokyo, Bunkyo, Tokyo 113-8656, Japan}

\author{Seungwu Han}
\email[]{hansw@snu.ac.kr}
\affiliation{Department of Materials Science and Engineering and Research Institute of Advanced Materials, Seoul National University, Seoul 08826, Korea}


\date{\today}



\maketitle


\renewcommand{\thefigure}{S\arabic{figure}}

\section{Supplemental Material}

\subsection{Neural network learning from sum}
\begin{figure}[h]
	\includegraphics[scale=1.2]{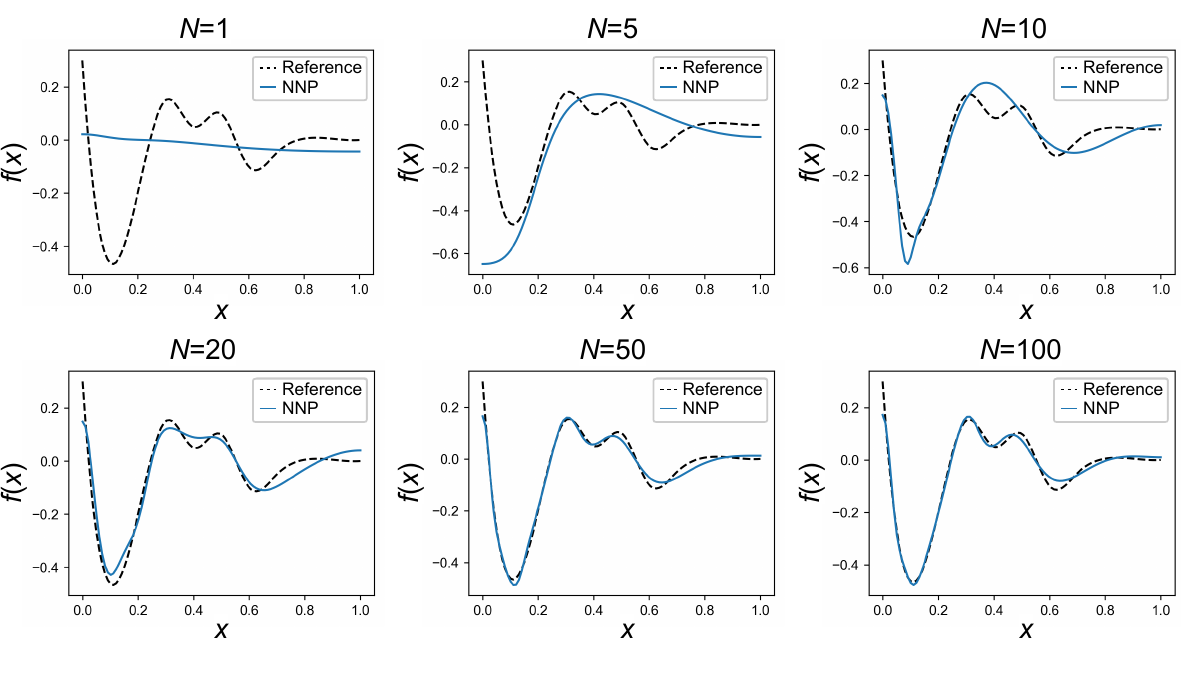}
	\caption{$f(x)$ predicted with trained NN with varying size of training set ($N$).\label{fig:s1}}
\end{figure}

We demonstrate with a simple mathematical model that the neural network (NN) is capable of inferring a target function  from  sums of the function values. We create a piecewise cubic spline $f(x)$ shown as the dashed line in Fig.~\ref{fig:s1}. The size of the training set ($N$) varies from 1 to 100, and each sample is a sum of $f(x)$ at five random $x$'s. A neural network of 8-30-30-1 structure is used, where $x$ is encoded into 8 $G_2$ symmetry functions (input nodes). Since the only sum of $f(x)$ is trained, NN cannot predict $f(x)$ at low $N$, but at sufficiently large $N$, NN can infer $f(x)$ accurately.

\subsection{Atomic energy mapping of Ni nanocluster}

\begin{figure}[h]
	\includegraphics[scale=1.8]{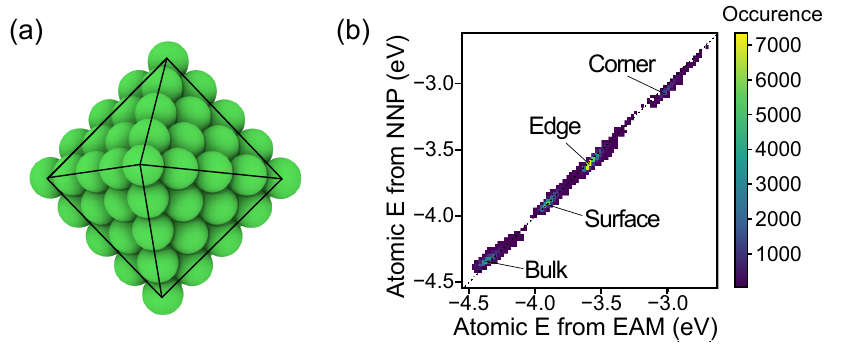}
	\caption{(a) Structure of Ni$_{85}$ cluster, which consist of 6 corner atoms, 36 edge atoms, 24 surface atoms, and 19 bulk atoms. (b) Correlation between atomic energy of EAM and NNP for Ni nanocluster.\label{fig:s2}}
\end{figure}

As a more practical and complicated example, we examine whether NN can identify underlying classical potential when only total energies are provided. Specifically, we train NNP on the total energy of embedded-atom-method (EAM) potential whose atomic energy is defined, and compare atomic energy of the two potentials. (The pairwise potential energy is split equally between the two atoms.) The training set consists of EAM MD snapshots of Ni$_{85}$ nanocluster, sampled in 10-fs interval. (See Fig.~\ref{fig:s2}(a).) After training, RMSE for total energy is 28.6 meV. Figure~\ref{fig:s2}(b) compares EAM atomic energy and NNP atomic energy.   Good correlations are observed with RMSE for the atomic energy of 25.1 meV, similar to that for the total energy. In particular,  NNP successfully resolves different configurations, namely corner, edge, surface, and bulk.

\subsection{Training set and computational details}
All DFT MD simulations are performed with Vienna ab initio simulation package (VASP)\cite{VASP1,VASP2,VASP3} based on the projector-augmented wave (PAW) pseudopotential. The generalized gradient approximation (GGA) in the form of Perdew-Burke-Ernzerhof (PBE) is used as exchange-correlation functional.

The training set in the first example consists of DFT MD snapshots of 64-atom fcc Si crystal at 1000 K in NVT and NPT condition (sampled in 20-fs interval). Cutoff energy of 300 eV and $\Gamma$-centered 2$\times$2$\times$2 {\bf k}-point grid is used, which ensures that the accuracy of DFT calculation is sufficient for the training procedure; the energy converges below 5 meV/atom and the forces converge to within 0.015 eV/$\text{\AA}$. NVT condition is imposed by scaling the velocities every 20 fs. NPT condition is applied with Langevin thermostat. Total dataset which consists of 500 structures (32000 atoms) is split into training set (70\%) and validation set (30\%) randomly.

The training set in the second example consists of MD snapshots of 128-atom Si(001) slab (2$\times$2 reconstructed) at 100, 300, 500, and 1000 K in NVT condition (sampled in 20 fs interval). Cutoff energy of 300 eV and $\Gamma$-centered 2$\times$2$\times$1  {\bf k}-point grid is used, and the forces converge below 0.022 eV/$\text{\AA}$. Total dataset of 500 structures (64000 atoms) are split into training set (70\%) and validation set (30\%) randomly.

The training set in the third example consists of DFT MD snapshots of 239-atom fcc Si crystal nano-cluster at 1000-1700 K in NVT condition (sampled in 10 fs interval at 1000-1500 K and in 20 fs interval at 1500-1700 K). Cutoff energy of 200 eV is used with only $\Gamma$-point sampling. NVT condition is imposed by scaling the velocities every 2 fs. Total dataset which consists of 1040 structures (248560 atoms) is split into training set (80\%) and validation set (20\%) randomly.

To represent local environment, we used atom-centered symmetry functions suggested by Behler\cite{GG}. The symmetry function vector $\mathbf{G}$ consists of 8 $G_2$ and 18 $G_4$ functions. The cutoff of 6.5 $\text{\AA}$ is used in the first and second examples, and 6.0 $\text{\AA}$ is used in the third example. Neural network with two hidden layers and 30 hidden nodes per layer is used (26-30-30-1 structure). Both total energy error and atomic force errors are minimized during the training process with L-BFGS optimizer.

\subsection{Prediction uncertainty of NNP}
Since the initial weights are randomized, different neural network is obtained between each training session.  Therefore, one NNP does not represent all NNPs and it is reasonable to estimate the prediction uncertainty of NNP in addition to the mean prediction value. We estimated the prediction uncertainty of NNP by training it five times with different initial conditions; the initial weights are randomized and the training set is also randomly selected from the total dataset each time. The uncertainty is, then, estimated as one standard deviation between five NNPs.

\subsection{The distance to the training set}

\begin{figure}[h]
	\includegraphics[scale=1.3]{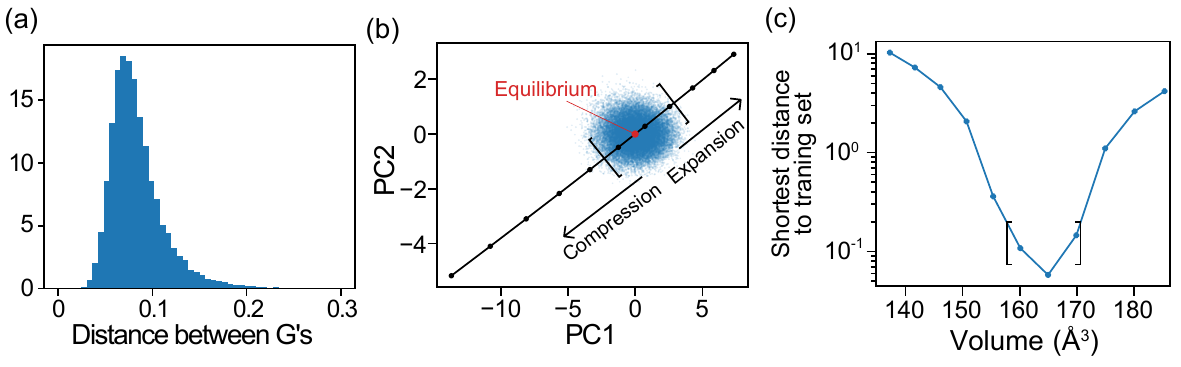}
	\caption{(a) The distribution of the distances among $\mathbf{G}$'s in the training set. (b) The distribution of $\mathbf{G}$ in the training set and EOS, projected onto principal component axes. (c) The shortest distance to $\mathbf{G}$ in the training set for each point in EOS. In both (b) and (c), square-bracket indicates the same range for EOS where the shortest distance to the $\mathbf{G}$'s in the training set is lower than 0.2.\label{fig:s3}}
\end{figure}

To show clearly that some of $\mathbf{G}$'s in the EOS is close to the training set, we plot in Fig. \ref{fig:s3}(a) the distribution of the distances among $\mathbf{G}$'s in the training set, and Fig. \ref{fig:s3}(b) the training points on the two axis from the principle component analysis. The solid disks in Fig. \ref{fig:s3}(b) correspond to $\mathbf{G}$ vectors for fcc Si along the equation of states (EOS). As can be seen in Fig. \ref{fig:s3}(a), most distances between $\mathbf{G}$'s in the training set are lower than 0.2. In Fig. \ref{fig:s3}(b,c), square-bracket indicates the range along EOS where the shortest distances to the training set is as low as the distances among $\mathbf{G}$'s in the training set (lower than 0.2). Therefore, $\mathbf{G}$'s in this range would be learnable.

\subsection{Metric to quantify connectivity in the $\mathbf{G}$ space}
 
\begin{figure}
	\includegraphics[width=12.9cm]{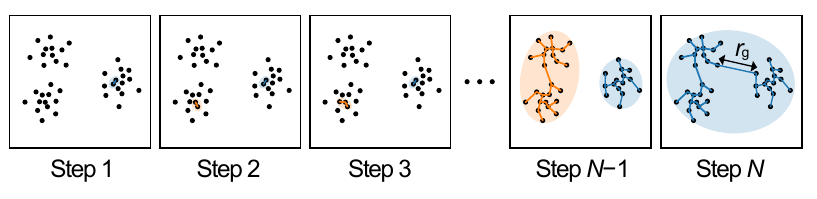}
	\caption{Schematic illustration of single-linkage clustering procedure to obtain $r_\text{g}$.\label{fig:s4}}
\end{figure}

\begin{figure}
	\includegraphics[scale=1.5]{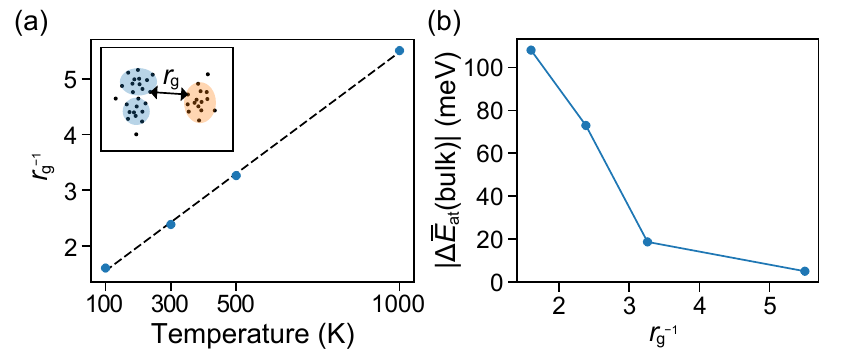}
	\caption{(a) $r_\text{g}^{-1}$ against the temperature of training set and (b) atomic-energy error (bulk) versus $r_\text{g}^{-1}$ for the second example (Si slab).\label{fig:s5}}
\end{figure}

For a systematic analysis of the second example in the manuscript, it would be useful to define a metric that measures the connectivity in the $\mathbf{G}$ space. To this end, we iteratively carry out single-linkage clustering of training points, a sort of hierarchical clustering used in the statistical analysis. At each step, two clusters with the shortest distance merge into one. (Initially, every training point represents an independent cluster.) The intercluster distance that reflects dissimilarity is set to the minimum Euclidean distance between two points from each cluster. The iteration proceeds until there remains only one large cluster with the size above a threshold value. We then define $r_\text{g}$ as the distance between the two lastly-linked large clusters. Here, the threshold is set to 0.5 times the number of structures in the training set, but the result is largely insensitive to a specific choice because the cluster size highly polarize near the end of iterations. Figure \ref{fig:s4} shows the schematic illustration of single-linkage clustering procedure to obtain $r_\text{g}$. Since $r_\text{g}$ approximates the maximum distance among distinct clusters [see inset of Fig. \ref{fig:s5}(a)], $r_\text{g}^{-1}$ can be regarded as the connectivity of the training set. Figure \ref{fig:s5}(a) shows $r_\text{g}^{-1}$ against the temperature of training sets for the slab model in Fig. 2 of manuscript. It is seen that $r_\text{g}^{-1}$ increases linearly with the temperature, supporting that the parameter correlates with the connectivity. Figure \ref{fig:s5}(b) plots $\Delta \bar{E}_\text{at}(\text{bulk})$ with respect to $r_\text{g}^{-1}$. It is seen that the mapping error is sufficiently low when $r_\text{g}^{-1}$ is larger than a certain cutoff ($\sim 4$). Although this cutoff value would vary with the system or training set,  $r_\text{g}$  would be useful in investigating the degree of connectivity in the training set quantitatively.


\bibliography{ref}